\pdfoutput=1

\documentclass[11pt]{article}

\usepackage[]{emnlp2021}

\usepackage{times}
\usepackage{latexsym}

\usepackage[T1]{fontenc}

\usepackage[utf8]{inputenc}

\usepackage{microtype}
\usepackage{multirow}

\usepackage{graphicx}
\usepackage{url}
\usepackage{booktabs} 
\usepackage{subfigure}
\usepackage{amsmath}
\usepackage{xcolor}

\newcounter{JSNumberOfComments}
\stepcounter{JSNumberOfComments}

\title{Predicting Anti-Asian Hateful Users on Twitter during COVID-19}

\author{
    Jisun An$^1$,
    Haewoon Kwak$^1$,
    Claire Seungeun Lee$^2$,
    Bogang Jun$^3$,
    Yong-Yeol Ahn$^4$\\
    
    $^1$School of Computing and Information Systems, Singapore Management University\\
    $^2$School of Criminology and Justice Studies, University of Massachusetts Lowell\\
    $^3$Department of Economics, Inha University\\
    $^4$School of Informatics, Computing, and Engineering, Indiana University, Bloomington\\
    
    \texttt{\{jisunan,hkwak\}@smu.edu.sg},
    \texttt{claire\_lee@uml.edu}\\
    \texttt{bogang.jun@inha.ac.kr},
    \texttt{yyahn@iu.edu}
}

\begin{document}
\maketitle
\begin{abstract}
We investigate predictors of anti-Asian hate among Twitter users throughout COVID-19. With the rise of xenophobia and polarization that has accompanied widespread social media usage in many nations, online hate has become a major social issue, attracting many researchers. Here, we apply natural language processing techniques to characterize social media users who began to post anti-Asian hate messages during COVID-19. We compare two user groups---those who posted anti-Asian slurs and those who did not---with respect to a rich set of features measured with data prior to COVID-19 and show that it is possible to predict who later publicly posted anti-Asian slurs. Our analysis of predictive features underlines the potential impact of news media and information sources that report on online hate and calls for further investigation into the role of polarized communication networks and news media. 
\end{abstract}


\section{Introduction}

The novel coronavirus pandemic (COVID-19) provides a unique opportunity to study the development of targeted racial animus at an unprecedented scale. 
Since the first known case of COVID-19 was reported in Wuhan, China~\cite{mallapaty2021after}, Asians have been a target of online and offline hate.
Multiple surveys have shown an increase in anti-Asian attitudes among Americans, which has manifested in xenophobic behaviors, that range from not visiting Asian restaurants or changing seats to avoiding Asians in public places~\cite{dhanani2020unexpected,croucher2020prejudice,reny2020xenophobia} to verbal and physical harassment~\cite{CSHE_hate_asian,AAPI_hate_asian}.

Online platforms, social media in particular, have been an exemplification, rather than an exception, of anti-Asian hate, hosting plenty of hateful content. 
Recent work has reported a striking increase of anti-Asian slurs on Twitter~\cite{ziems2020racism} and negativity towards China on Twitter and Reddit~\cite{schild2020go}. 
Many Asians have expressed an increased level of anxiety and depression due to the recent development of racial animus~\cite{doi:10.1080/01419870.2020.1851739}.
As an attempt to investigate online hate towards Asians during COVID-19, several studies have introduced new datasets and methods for hate detection towards Asians~\cite{ziems2020racism,vidgen2020detectingasian}.

While most studies focus on the \emph{content}---e.g., whether a given text is a hate speech or not, \emph{user}-level analysis---e.g., whether a given user will post hateful content or not---is surprisingly under-explored, although this question can bring useful and important insights.
First of all, considering the fact that only a handful of users produce the vast majority of  misinformation~\cite{ccdh2021antivaxxers} and hateful content (\S\ref{sec:shared_news_media}), focusing on users may be an efficient way to tackle online hate. For instance, identifying `influential' users who tend to produce a large volume of such content and have the capacity to be the center of the discussion can lead to a better intervention than identifying individual instances of hate speech.
Second, understanding the risk factors for hate speech may also provide an opportunity to nudge the users and keep them from being radicalized. This pathway is only possible when we examine individual-level risk factors.  
Finally, because there tends to be much more data available about individual users than individual tweets or posts~\cite{qian-etal-2018-leveraging}, we may be able to develop a more accurate model as well as in-depth insights into the development pathways of online hate. 
At the same time, we emphasize that care must be taken when taking user-based approaches because the prediction of future offenses and misbehaviors can potentially lead to algorithmic bias as shown in the case of recidivism prediction~\cite{angwin2016machine}. 
Thus, our work should be considered an early step towards understanding online hate, and we caution that the translation of our results into practice would require much more nuanced investigation and decision-making.


In this paper, we tackle online hate towards Asians by focusing on users, particularly focusing on the risk indicators of hate towards Asians.
We examine users' language use in their tweets and information sources that they follow on social media. 
As polarization and echo chamber have been commonly observed on social media~\cite{garimella2017long,an2019political}, it is reasonable to expect that users would choose to whom they listen, often in the way to strengthen their biases. Thus we expect to see signals from the identities of the followings of users. 
Although our retrospective case-control design does not allow strong causal inference (i.e., we cannot discern whether a user is affected by whom they follow or their following is a manifestation of their existing bias), our study may shed light on potential mechanisms and pathways towards online hate. 
Also, while our work focuses on online hate, it would  provide helpful insights for modeling offline hate crimes as well~\cite{Relia_Li_Cook_Chunara_2019}.

Our contributions can be summarized as two-fold. First, we analyze user-level predictors of hate speech towards Asians. We study the impact of both linguistic and information sources, namely \textit{(i) Content features} (social media posts published before COVID-19) and \textit{(ii) Content-agnostic features} (what kind of information a user is exposed to, how a user interacts with the platform and other users, etc). We further study the level of expressed hate. Second, we will release our dataset (features) and the code necessary to replicate our results\footnote{The data and the corresponding code are available at \url{https://github.com/anjisun221/Anti-Asian-Slurs}}. 


\section{Related Work}


Racial animus has been widely studied from the perspective of its impact on individuals' health~\citep{sellers2003racial}, 
economic development~\citep{card2008tipping}, voting~\citep{stephens2014cost} and social unrest~\citep{BBC2020georgefloyd}.
Understanding how racial animus is developed is essential to prevent the risk of further intensifying racial animus and reduce its harm to society.

Online hater has been studied actively in recent years. On internet forum~\cite{de-gibert-etal-2018-hate}, Wikipedia~\cite{wulczyn2017ex}, News media comments~\cite{10.1111/jcom.12104}, Twitter~\cite{davidson2017automated,waseem2016hateful,waseem2016you,founta2018large}, YouTube~\cite{Salminen2018}, online games~\cite{kwak2015exploring}, and many other online platforms~\cite{zannettou2018gab}, 
different types of online hate have been investigated. 
While leveraging language to detect hate speech is not new~\cite{schmidt2017survey}, user-level modeling for detecting a future (event-driven) risk of posting hate speeches is relatively unexplored.
Among a few,
\citet{almerekhi2020these} identify linguistic markers that trigger toxicity in online discussions, 
\citet{ribeiro2018characterizing} detect the users who express online hate by checking their profiles or linguistic characteristics,  \citet{qian-etal-2018-leveraging} incorporate users' historical tweets and user similarities for hate speech detection, and \citet{lyu2020sense} compares hate instigators, targets, and general Twitter users by self-presentation, Twitter visibility, and personality trait. \citet{lyu2020sense} characterize Twitter users who use controversial terms associated with COVID-19 (e.g., ``Chinese Virus'') and those who use non-controversial terms (e.g.,  ``Corona Virus'') in terms of demographics, profile features, political following, and geo-locations.
Unlike previous studies, our work focuses on not only user features but also on the content posted \textit{before} they start to use Asian slurs, studying the development of anti-Asian hate attitudes.

Hate towards Asians during COVID-19 has appeared through diverse forms. For example, anti-Asian slurs are increasingly used~\cite{schild2020go}. 
Also, over 40\% of survey respondents in the U.S. would engage in discriminatory behavior towards Asians due to fear of COVID-19, lack of knowledge about the virus and trust in science, and more trust in Donald Trump~\citep{dhanani2020unexpected}. 
\citet{reny2020xenophobia} reported that xenophobic behavior as well as concerns about the virus is associated with anti-Asian attitudes.

The context and role of social media in online and offline hate have been argued to be important.
\citet{croucher2020prejudice} examined a link between social media use and xenophobia toward Asians. 
\citet{ziems2020racism} demonstrated that they could identify hate and counterhate tweets with an AUROC of 0.85. 
\citet{vidgen2020detectingasian} built an annotated corpus for four Asian prejudice-related classes, including criticism without being abusive. Their best model achieved a macro-F1 score of 0.83, but the only 59.3\% of hostility tweets were correctly identified.

\section{Data}

We first identify those who express hate towards Asians by collecting data from Twitter in three steps: 1) compile a list of anti-Asian slurs; 2) collect tweets with any of the anti-Asian slurs; and 3) identify those who have used anti-Asian slurs after the pandemic began (we call them \textit{hateful users})  and collect their historical tweets from June 8, 2019 to May 8, 2020. Then, as a control set, we randomly sample users who have tweets about COVID-19 (we call them \textit{reference users}) and collect their historical tweets during the same period. 
%
%

\paragraph{Anti-Asian slurs.} 
We compile a list of anti-Asian slurs by combining 1) Wikipedia's list of ethnic slur words~\cite{wikiasianslur} that includes `chink,' `chinazi,' and `chicom,' and 2) a set of COVID-19 specific anti-Asian slurs, such as `wuflu' and `kungflu'~\cite{schild2020go}. 
The full list of 33 anti-Asian slurs is in 
Appendix~\ref{sec:list_asian_slur_words}.

\paragraph{Hateful users.} 
Using \texttt{Twint}~\cite{poldi2021twint}, in May 2020, we collect tweets that contain any of the anti-Asian slurs from December 31, 2019 to May 1, 2020, resulting in 190,927 tweets posted by 120,690 users. 
We then consider those who 1) live in the U.S. and 2)  posted at least two tweets with anti-Asian slurs. We use  self-declared locations in user profiles to infer their state-level location and exclude users without identified locations. This leaves us with 3,119 \emph{hateful users}.

\paragraph{Reference users.} As a control set, we construct a set of non-hate users. Using Twitter's Streaming API, we collect 250M tweets that include COVID-19 related keywords (e.g., ``covid'' and  ``coronavirus'') from January 13 to April 12, 2020. The full list of COVID-19 keywords used for this data collection can be found in Appendix~\ref{sec:list_covid19_words}. 
From this dataset, we randomly select 3,119 \emph{reference users}, whose location can be detected at state-level and who have not used any anti-Asian slurs.

\paragraph{Historical tweet collections.} 
For the two user groups,
we collect their historical tweets posted for 11 months from June 8, 2019 to May 8, 2020 using \texttt{Twint} package~\cite{poldi2021twint}.  
In total, we collect 18,952,895 tweets where 15.91M (83.94\%) tweets of hateful users and 3.04M (16.06\%) tweets of reference users. 
We find 15,728 tweets (0.00083\%) containing anti-Asian slurs. 

\paragraph{Pre- and Post-COVID-19 tweets}
We use December 31, 2019 when China confirmed the first COVID-19 case as the start date of COVID-19. 

\paragraph{Refining reference users.} 
As explained above, our reference users are randomly selected from a large collection of COVID-19-related tweets. We find that 19 reference users (0.6\%) have used anti-Asian slurs before COVID-19. We exclude those users for the rest of the study.


\paragraph{Refining hateful users.}
Some hateful users have used anti-Asian slurs before COVID-19.
Through manual examination, we notice that most of them are activists in Hong Kong, expressing negativity through slur words targeting China. We exclude these users from further analyses.

By contrast, the use of slur words like ``wuflu'' and ``kung flu'' began to increase starting from March 9, 2020, a day when Italy extends emergency measures nationwide~\cite{guardian2020coronavirus}, and showed a sharp spike on March 16, 2020, when the former US president Donald Trump referred to COVID-19 as ``Chinese Virus'' on Twitter~\cite{yam2020trump}. 
Since then, all anti-Asian slurs becomes widely used. 
We focus on these users who turned hateful after the COVID-19. 

\paragraph{Low- vs High-level Hateful users.}
We further divide hateful users into two groups based on the level of expressed haterism towards Asians. 
In our data collection, there are 8,769 tweets that contain at least one slur word. The average number of tweets with a slur word per user is four (min: 1, median: 3, max: 126). 
Thus, we further divide users into two groups based on the average tweets with slurs: \emph{Low-level hateful users} 
with less than or equal to three tweets with anti-Asian slurs and \emph{High-level hateful users} 
with at least four tweets with anti-Asian slurs.

\begin{table}[t]
\centering \small
\begin{tabular}{l|r|rrr|r}
\toprule
& \multicolumn{1}{c|}{\textbf{Ref}}  & \bf Hate &  Low H & High H & \multicolumn{1}{c}{\textbf{Total}} \\
\midrule
\bf Users & 2,443 & 1,899 & 1,316 & 583  & 4,342\\
\bf Tweets & 250k & 8.2M & 4.8M & 3.4M & 10.1M\\
\bottomrule
\end{tabular}
\caption{Dataset overview}\label{tab:dataset}
\end{table}


\paragraph{Bot detection.} 
We further remove `bot' accounts to better capture genuine human behaviors. We use Botometer API~\cite{davis2016botornot}, a popular supervised machine learning tool that checks Twitter accounts for possibly automated activity by using features including content, network structure, and profile.
Given a Twitter account, Botometer API returns a score from 0 to 1, where 1 is the most likely to be a bot. 
We use the fifty-percent threshold (Botometer score = 0.5), which has proven effective in prior studies~\cite{davis2016botornot}, to remove potential bot accounts. 

\paragraph{Data summary.} 
As a result, our data collection for further analyses contains 8,201,510 and 2,498,246 tweets posted by 1,899 hateful users and 2,443 reference users, respectively. We also collect their network information---whom they follow by using Twitter REST API. Details are in Table~\ref{tab:dataset}.

\section{Comparative Exploration of Hateful and Reference Users}

Prior to predicting future hate behaviors of users, we first conduct a comparative exploration between hateful users and reference users to better understand their discriminating characteristics.

\subsection{How much and what they write}
\subsubsection{Twitter Activity} 
The first noticeable difference between hateful users and reference users is their activity level in terms of the number of written tweets after COVID-19. 
On average, both user groups increased their activities, and the increase is more considerable for hateful users. 
We evaluate the statistical significance by 1,000 bootstrapped samples. The bootstrapped average percent increase among hateful users is much higher than reference users:
$90.12\,\%$ ($\pm 0.419\,\%$, 95\% CI) vs. $24.52\,\%$ ($\pm 0.107\,\%$, 95\% CI). 
Furthermore, we observe that the increase in activity after COVID-19 is driven more by those High-level hateful users than Low-level ones.  The bootstrapped average percent increase among High- and Low-level hateful users are $118.97\,\%$ ($\pm 0.977\,\%$, 95\% CI) and $77.22\,\%$ ($\pm 0.381\,\%$, 95\% CI), respectively.


\subsubsection{Representative Words of User Groups}

As an exploratory analysis, we identify the representative words of user groups by using the log-odds ratio proposed in \citep{monroe2008fightin}. 
For each group, we aggregate all their pre-COVID-19 and post-COVID-19 tweets separately, creating the four corpora. 
To remove rare jargon, we eliminate terms that appear less than 10 times. We then extract all unigrams and compute their log-odds ratio. 
As the prior, we compute background word frequency on two separate random Twitter datasets for pre-COVID-19 and post-COVID-19, sampled from Twitter Decahose data.  
The unigrams are then ranked by their estimated $z$-scores. 

\begin{figure}[h!]
\centering
\subfigure[(Pre) Hateful users ]{\includegraphics[width=.49\columnwidth]{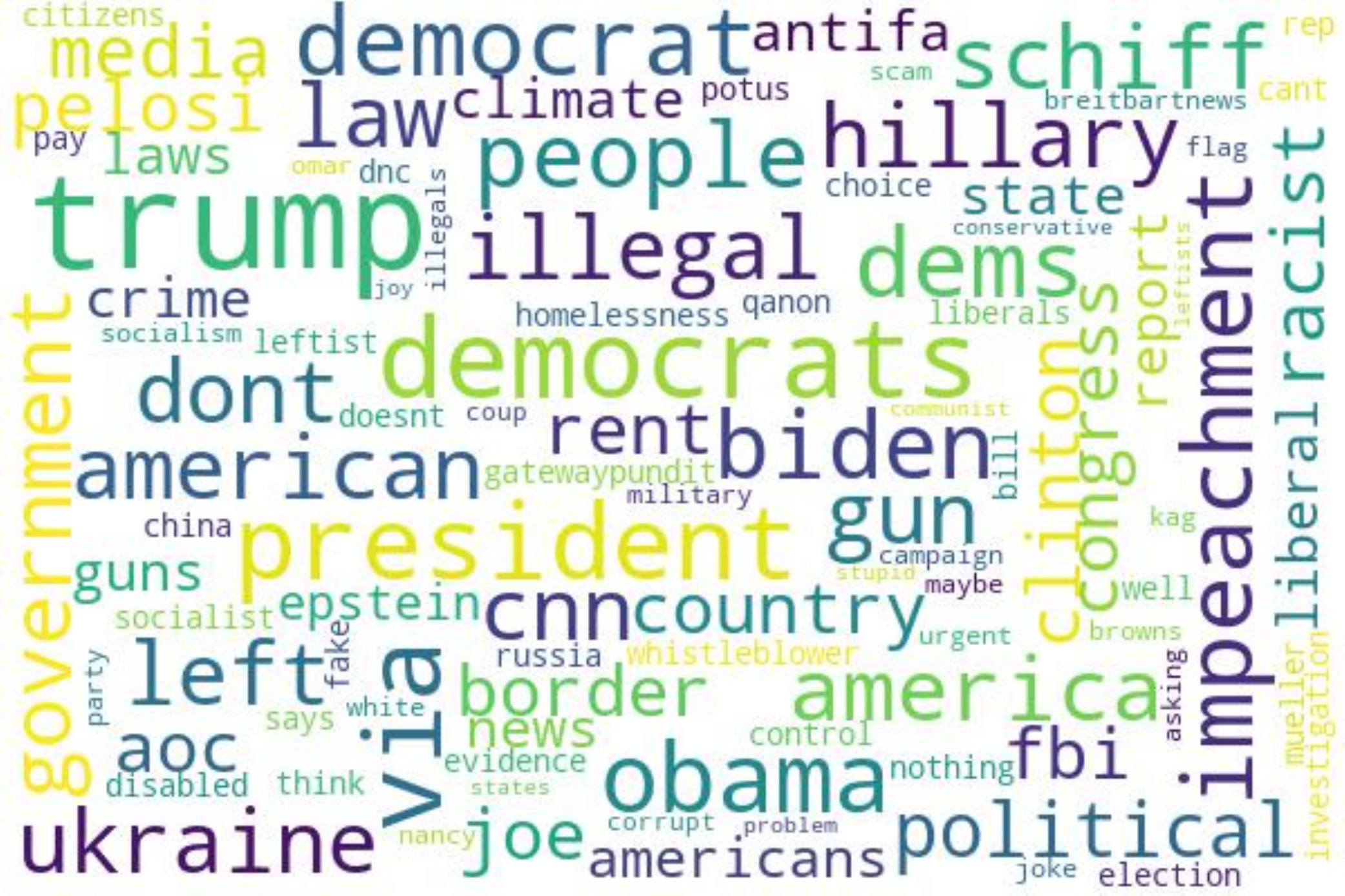}\label{fig:hate_precovid}}
\subfigure[(Pre) Reference users]{\includegraphics[width=.49\columnwidth]{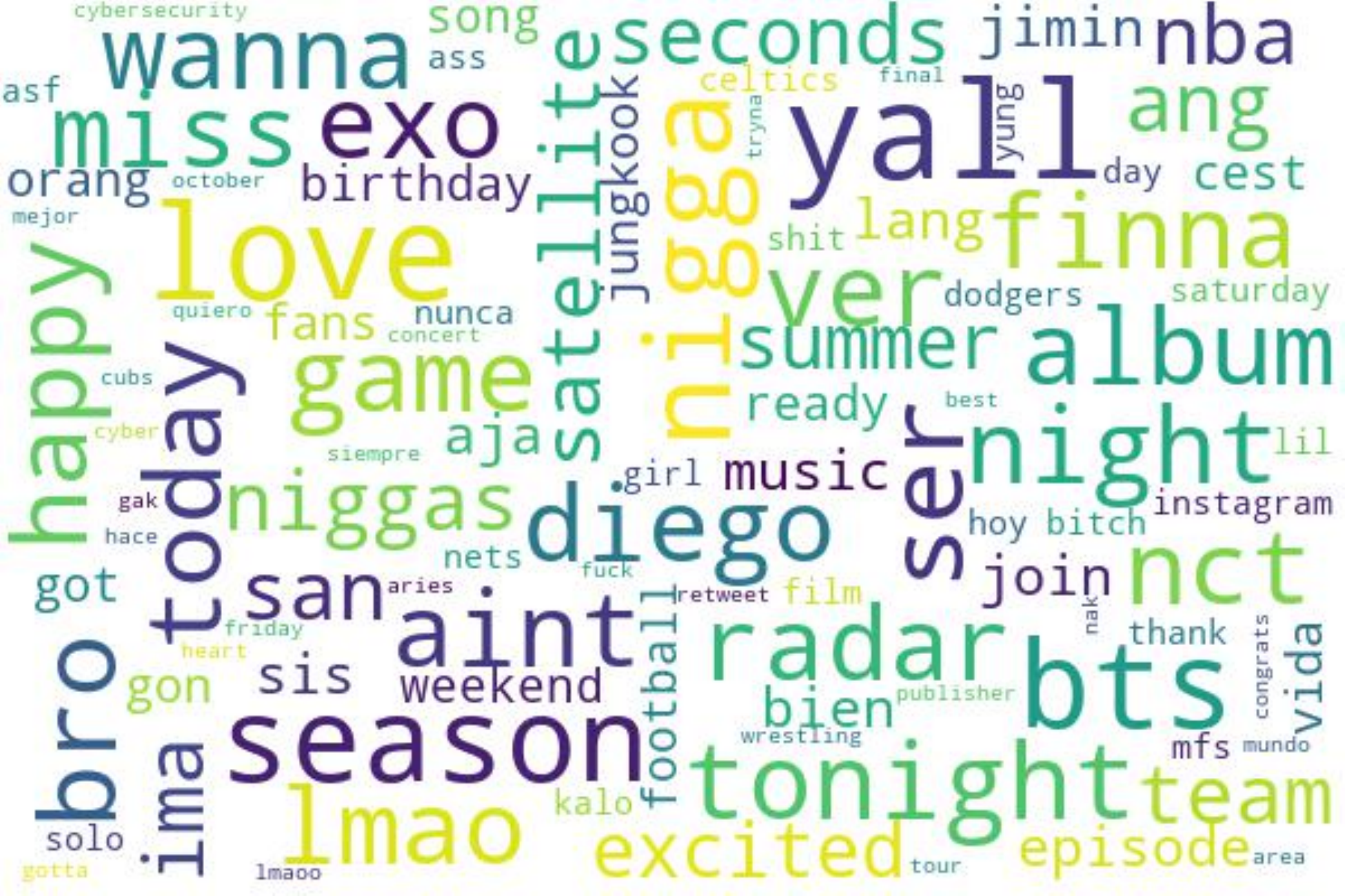}\label{fig:ref_precovid}}
\subfigure[(Post) Hateful users ]{\includegraphics[width=.49\columnwidth]{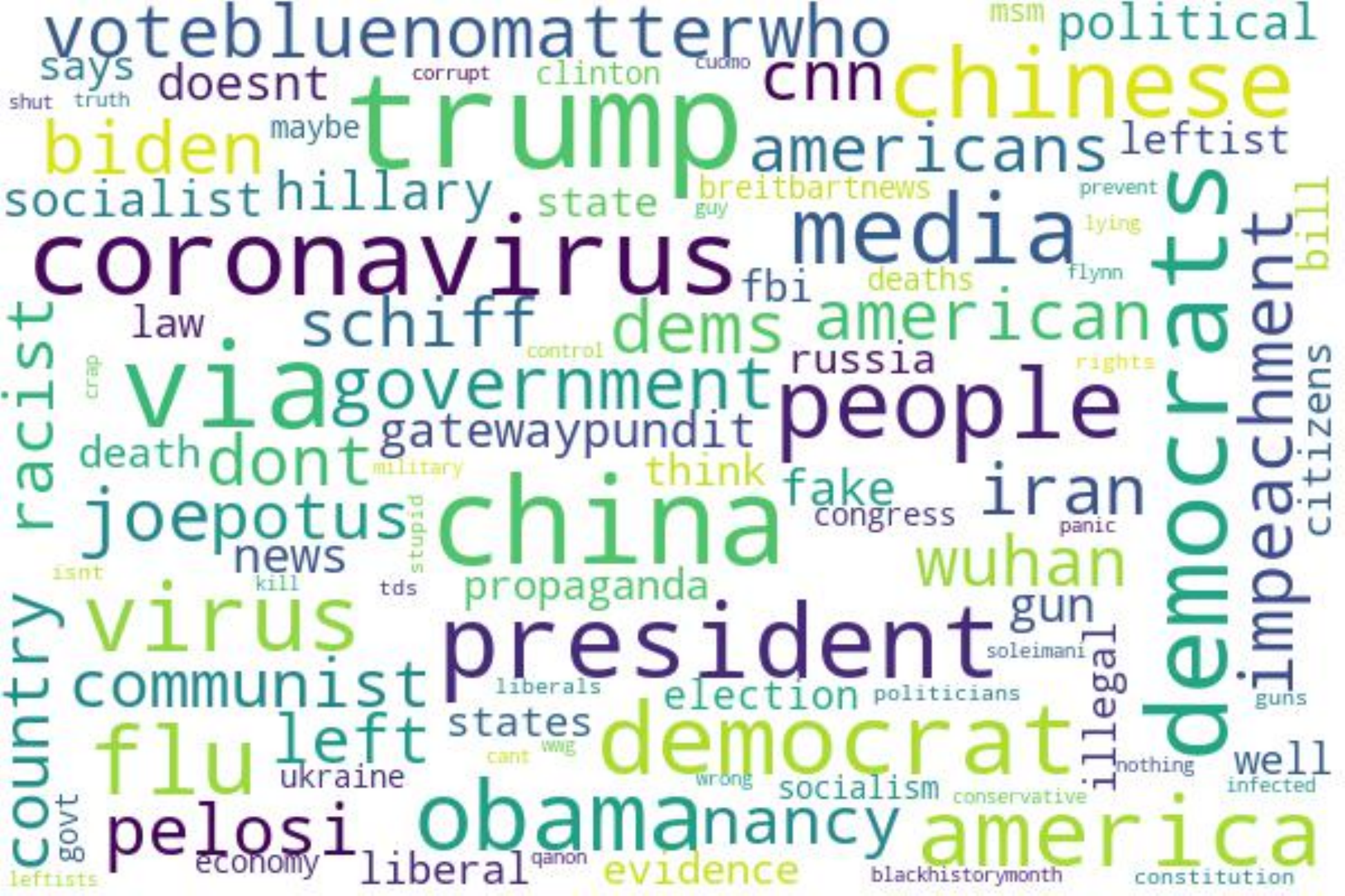}\label{fig:hate_postcovid}}
\subfigure[(Post) Reference users]{\includegraphics[width=.49\columnwidth]{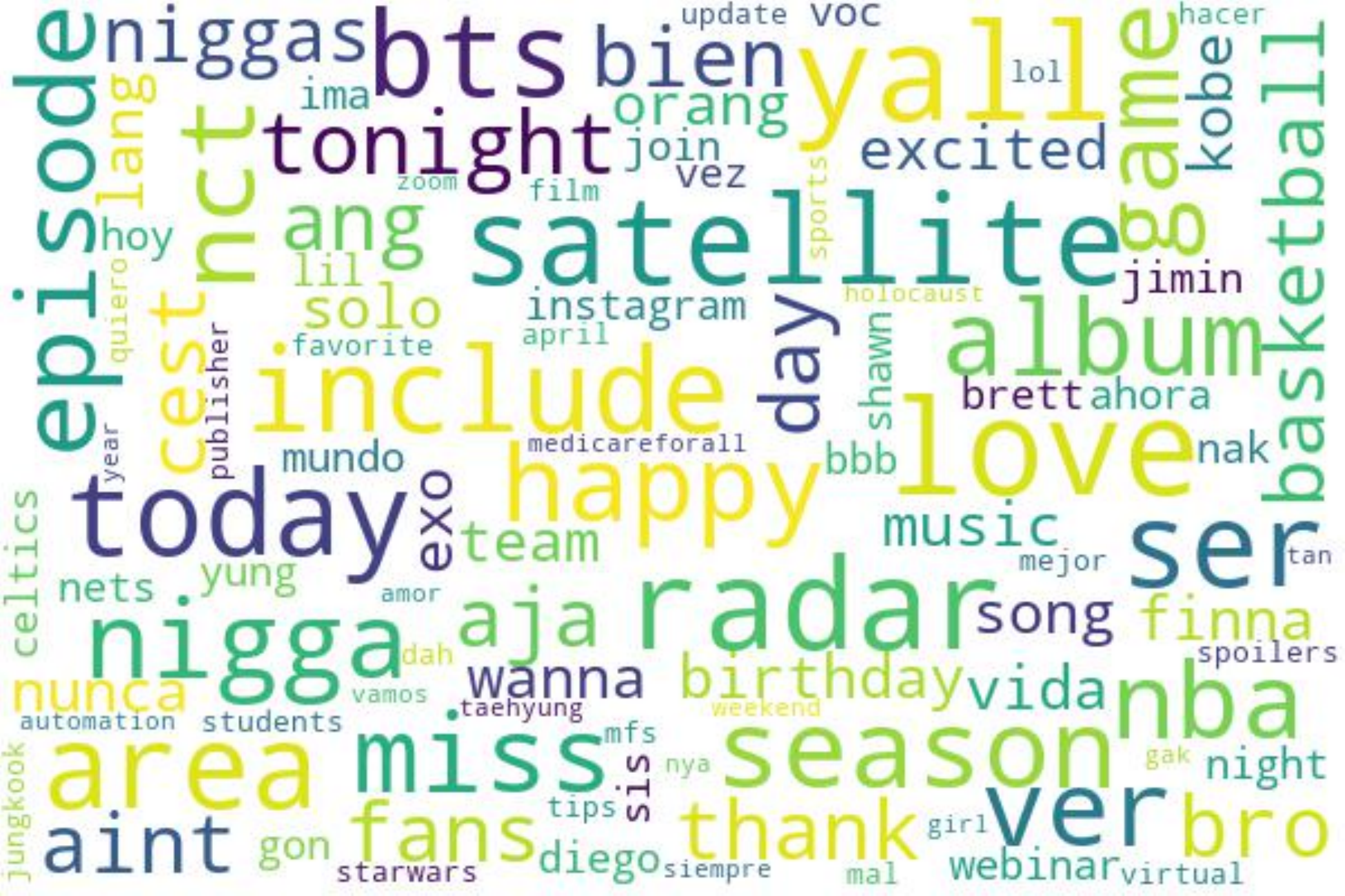}\label{fig:ref_postcovid}}
\caption{Over-represented unigrams of Hateful users and Reference users for Pre- and Post-COVID-19}
\label{fig:wordcloud}
\end{figure}

Figure~\ref{fig:wordcloud} shows the 100 most over-represented unigrams from each corpus.
For hateful users before COVID-19 (Figure~\ref{fig:hate_precovid}), US politics-related words (e.g., obama, biden, clinton, conservative, and democrat) and international political issue-related words (e.g., iran, border, war, and  ukraine) are presented.
Some words indicate that these users are likely to be right-wing: 1) maga (``Make America Great Again'') is a campaign slogan used in American politics popularized by Donald Trump, 2) terms like leftists and socialist, and 3) right-wing news media, including breitbartnews, gatewaypundit, and dailycaller.
Another interesting word set contains names: soros (likely George Soros), omar (likely Ilhan Omar), schiff (likely Adam Schiff), and nancy (likely Nancy Pelosi), who opposed Donald Trump. 
Also, a hashtag `\#votebluenomatterwho' seems to be used by hateful users. 
        
The reference users' prominent words before COVID-19  (Figure~\ref{fig:ref_precovid}) look more casual than hateful users, which also show the validity of our reference sample to some extent. 
The over-represented words are related to sports and entertainment (e.g., nba, football, game, and music), and K-pop (e.g., bts, nct, and exo), which is one of the globally popular topics on Twitter~\cite{kim2021kpop}. 
Also, words with positive connotations (e.g., love, happy, thank, congrats, and excited, lmao (``laughing my ass off'')) are appeared. We find some N-word expressions from this group, suggesting the presence of more black Twitter users in this group.


After COVID-19, hateful users seem to actively write tweets about COVID-19 (Figure~\ref{fig:hate_postcovid}), including China, Chinese, virus, flu, death, and Wuhan. 
Words related to infodemic, such as propaganda, fake, wrong, lying, and truth, are also presented, which is well aligned with a previous report~\citep{cinelli2020covid}.

Lastly, Figure~\ref{fig:ref_postcovid} shows reference users' over-represented words after COVID-19. We do not see many changes for this group but note two particular words: kobe and medicareforall. Kobe (Bryant) is an NBA superstar who passed away in January 2020. Since `medicareforall' is one of the key slogans of the Democrats, this suggests that more left-wing users may be present in the reference users.

\subsubsection{How their tweets are engaged by others}

We examine how actively other users engage in hateful users' tweets. 
The average retweet (like) counts of hateful users is 1.88 (7.88),  while that of reference users is 1.47 (7.68). 
We run a Mann-Whitney's $U$ test to evaluate the difference and find a significant effect of group for retweet counts ($U$ = 6.05,  $p <$ 0.001), but not for like counts. 
Furthermore, high-level hateful users tend to have higher retweet counts but lower like counts than low-level users.
The average retweet (like) counts are 2.97 (1.01) and 1.11 (6.29) for high-level and low-level hateful users, respectively. The differences of retweet and like counts are statistically significant ($p <$ 0.001 and $p <$ 0.05, respectively).  

Lastly, tweets with anti-Asian slurs tend to be more retweeted and liked than other tweets of hateful users. The average retweet (like) counts are 4.52 (10.87) for tweets with anti-Asian slurs and 1.87 (7.88) for other tweets. We find that the like count difference is statistically significant ($p < 0.001$) but not the retweet count. 
The results indicate that the tweets posted by hateful users are more likely to be propagated and thus get exposed to the target group. Moreover, it suggests that there may exist positive feedback for hateful tweets---expressing hate tends to increase engagement, which in turn may nudge users to post more hate tweets.


\subsection{What They Consume and Share}

\subsubsection{Shared News Media}\label{sec:shared_news_media}

News is known to be powerful in shaping people's opinions and behaviors~\cite{mccombs2020setting}. A potential bias or factuality of news reporting thus can influence one's attitude towards Asians. 
To examine what information users are exposed to, we opt for analyzing news URLs shared by users. In doing so, we use media-level factuality ratings on a 7-point scale (Questionable-Source, Very-Low, Low, Mixed, Mostly-Factual, High, and Very-High) and bias ratings on a 7-point scale (Extreme-Left, Left, Center-Left, Center, Center-Right, Right, and Extreme-Right) annotated by the Media Bias/Fact Check (MBFC)~\cite{mbfc}.

\begin{figure}[h!]
\centering
\subfigure[Bias ]{\includegraphics[width=.49\columnwidth]{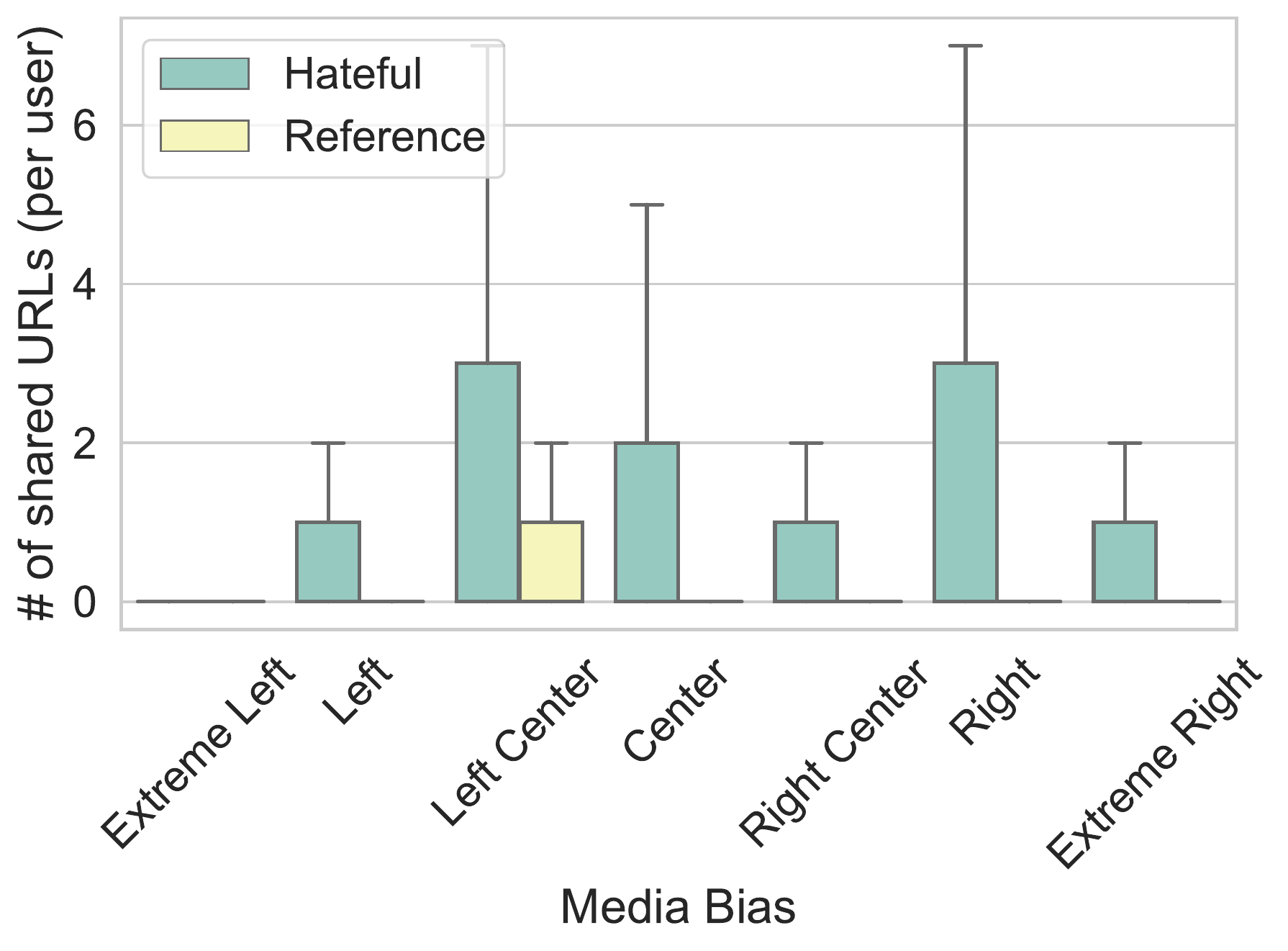}\label{fig:media_bias}}
\subfigure[Factuality]{\includegraphics[width=.49\columnwidth]{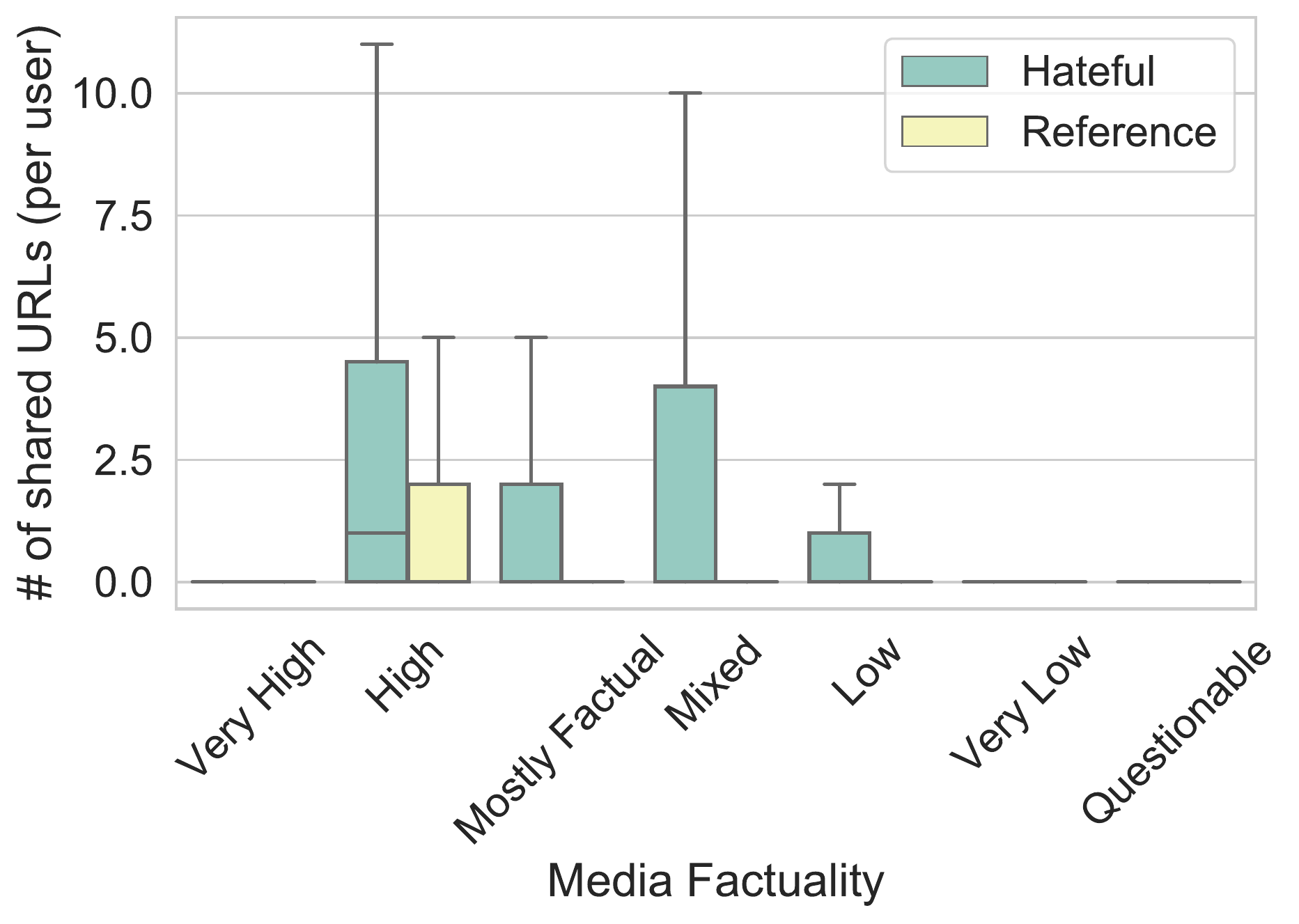}\label{fig:media_factual}}
\subfigure[Bias, High vs Low ]{\includegraphics[width=.49\columnwidth]{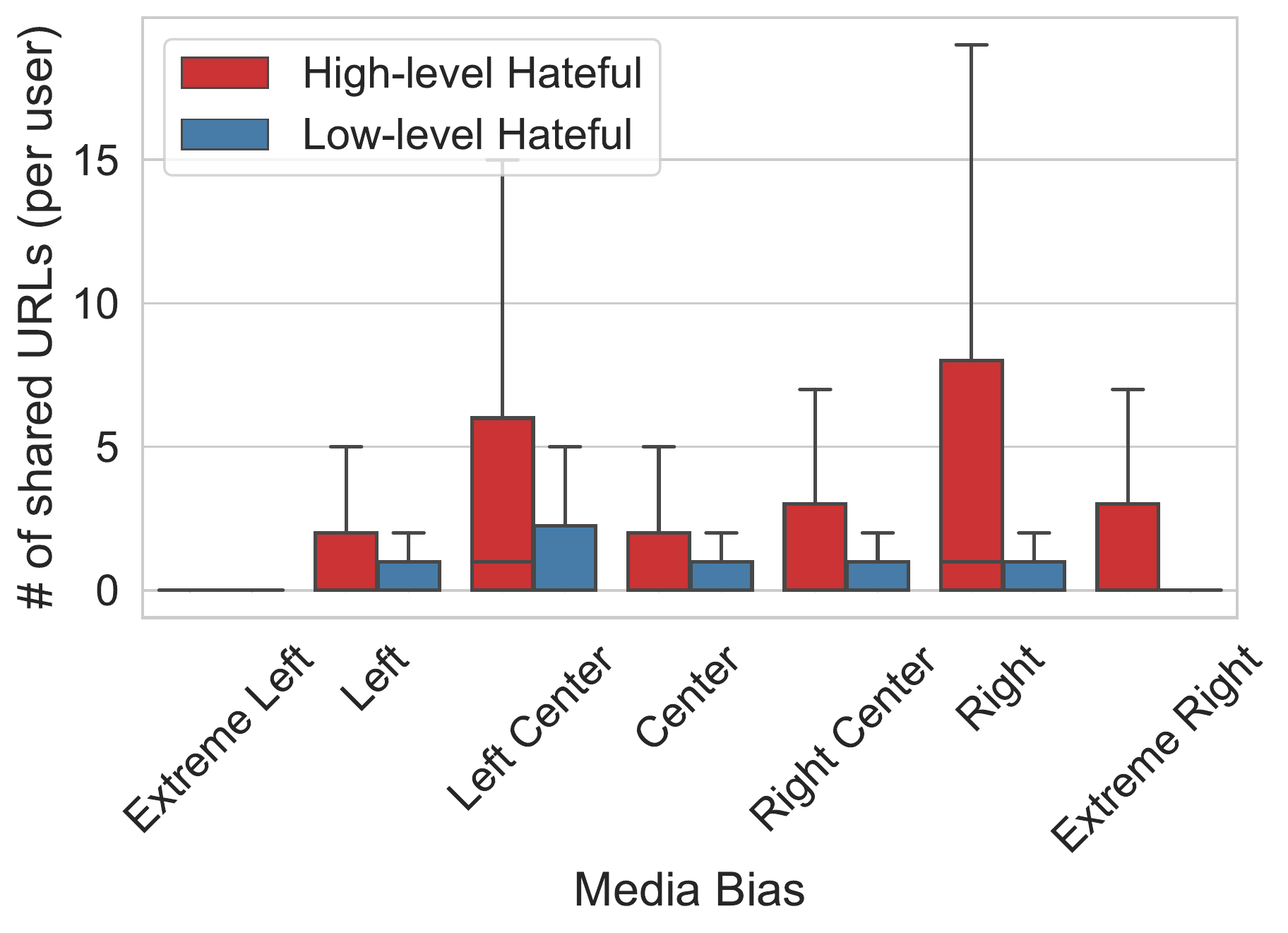}\label{fig:media_bias_level}}
\subfigure[Factuality, High vs Low]{\includegraphics[width=.49\columnwidth]{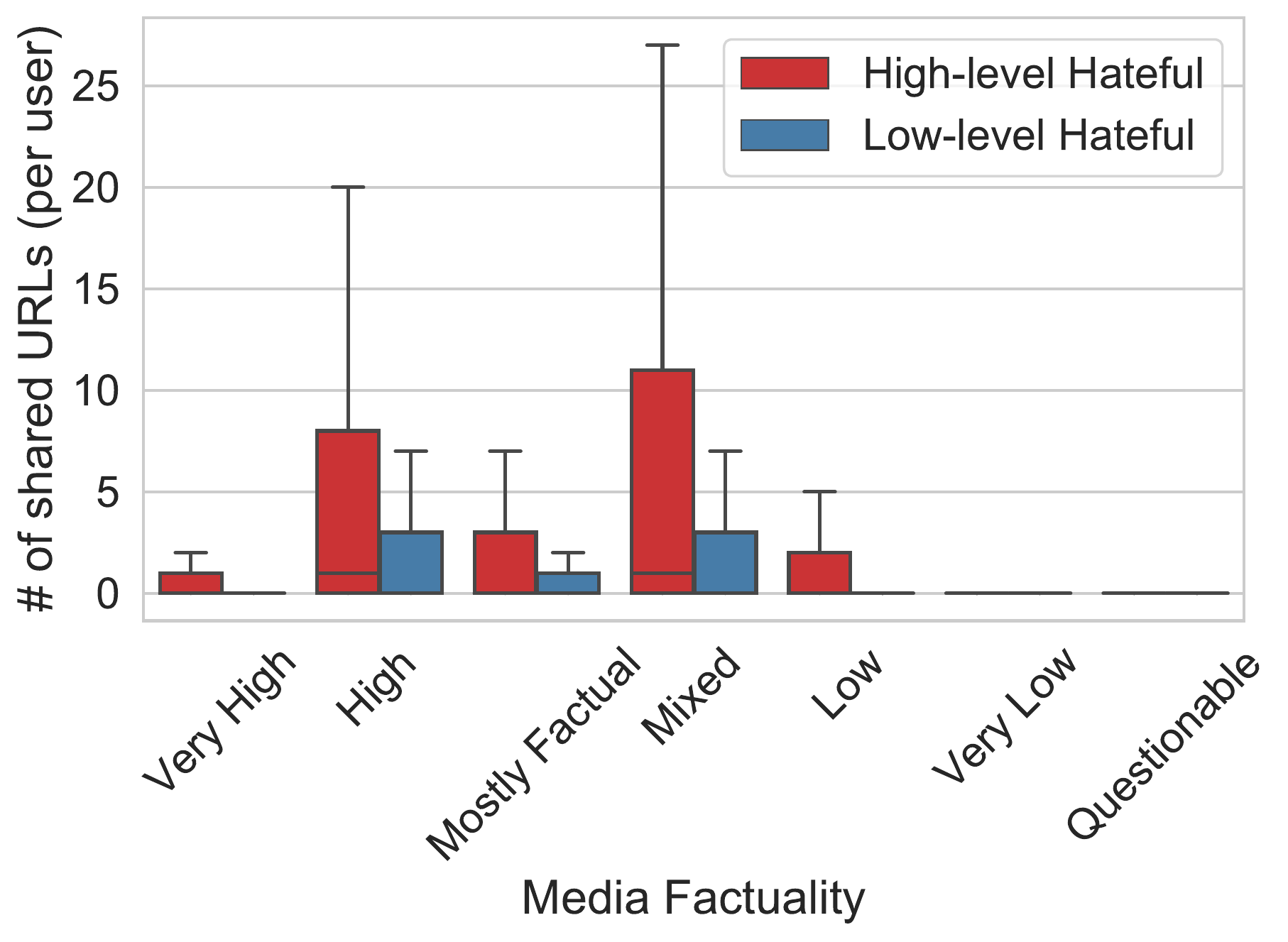}\label{fig:media_factual_level}}
\caption{Shared News Media before COVID-19}
\label{fig:shared_news_media}
\end{figure}

We compare the average number of shared URLs for each of categories in media bias (Figure~\ref{fig:media_bias}) and factuality (Figure~\ref{fig:media_factual}) between hateful and reference users from their pre-COVID-19 tweets.
Across all categories, hateful users share more news URLs than reference users. 
While the representative word analysis hints that hateful users are more likely to be right-wing, their shared news are fairly diversified. 
While hateful users share URLs from credible news media, they also share many news URLs from less credible news media. 
When comparing high-level and low-level hateful users, active news sharing behavior of hateful users mostly comes from high-level hateful users.

\subsubsection{Followings}

The accounts following is yet another proxy of information sources. We examine Twitter accounts that are the most followed by each group. 
For reference users, the top 5 are `BarackObama', `realDonaldTrump', `nytimes', `AOC', and `cnnbrk', while those by hateful users are 
`realDonaldTrump', `RealJamesWoods', `POTUS', `DonaldJTrumpJr', and `TuckerCarlson'. 
In the top 50 Twitter accounts followed by the two groups, only 5 are in common, indicating that the information sources of the two groups are distinctive. 
Among hateful users, the overlap between low- and high-level hateful user groups is high---45 of the top 50 are shared.


\subsection{Summary}

In sum, hateful users, as a group, exhibit noticeable differences in comparison with  the reference users, including 1) being more active on Twitter during COVID-19, 2) using more words about politics (especially right-wing), 3) sharing significantly more URLs of news media, and 4) having distinctive information sources. 

High-level hateful users can be further differentiated from low-level ones by the following features: 1) they increased Twitter activity even more (about 54\% more than low-level hateful users) after COVID-19, 2) their tweets tend to be more shared and liked, and 3) they share more URLs published by news media of extreme bias and low factuality. 




\section{Predicting Hateful Users}

Our comparative exploratory analysis indicates that there exist significant differences between these groups, raising an important question: can we \emph{predict} those who turned hateful by only using the information available prior to the pandemic? The possibility of such prediction, combined with the analysis of prominent predictors, may help us understand the pathways into xenophobic haterism and design potential interventions. 

\subsection{Features}
\label{sec:features}

For each user, we extract 1) content features and 2) content-agnostic features for training a classifier to predict the future hate expression towards Asians. 

\subsubsection{Content Features}
From the over-represented words in Figure~\ref{fig:wordcloud}, we observe that word usage is distinct between the two user groups. 
We thus extract content features from the pre-COVID-19 tweets of every user. 
Since the distribution of the number of tweets per user is skewed, we sample 198 tweets per user, a median number of pre-COVID-19 tweets per user, and extract content features from the sampled 198 tweets.

\paragraph{NELA Features:} 
We extract following linguistic features by using the News Landscape (NELA) toolkit~\cite{horne2018assessing}: text structure (e.g., POS tags), sentiment, subjectivity, complexity (e.g., readability), 
bias~\citep{recasens2013linguistic,mukherjee2015leveraging}, 
and morality~\cite{graham2009liberals,lin2018acquiring}.
These features are known to be indicative for detecting fake information and the political bias of news sources. 
We expect that they help to capture aspects of factuality or political bias of hateful users' tweets. 
For each user, we obtain her NELA features by averaging the NELA features of her tweets. 

\paragraph{Psycholinguistics Features:} We extract features that relate to emotional and psychological characteristics of users by using Linguistic Inquiry and Word Count (LIWC)~\cite{pennebaker2015development}. 

\paragraph{Embedding Features:} 
We encode each tweet using Sentence BERT (SBERT)~\cite{reimers2019sbert} for the following reasons: 1) the distribution of the number of tweets per user is highly skewed, and 2) tweet has a sentence-like structure and length. 
Thus, we opt for SBERT and average the SBERT representations across the sample tweets to obtain a user-level representation. 
The averaged user-level representation is a 768-dimensional vector.

\paragraph{Shared News Media:} 
We also consider news media shared by users as features because they can be a proxy for information sources that users consume and propagate. Using seven categories of bias and factuality of news media introduced in \S\ref{sec:shared_news_media}.

\subsubsection{Content-agnostic features}
We model users based on how they interact with others and how they portray themselves to their audiences by using their Twitter profiles. 

\paragraph{Twitter Statistics:} We use basic information extracted from a Twitter profile, such as 1) whether a user's account is verified by Twitter; 2) the number of days since the account was created; and 3) four Twitter statistics including the number of followers, followings, tweets, and favorites.

\paragraph{Twitter Profile Description:} Since most of the Twitter profile description (bio) is short and has a sentence-like structure, we encoded the user's profile description using SBERT, like \citet{baly-etal-2020-written} did for obtaining followers' representations. 

\paragraph{Twitter Following:} We consider those who followed by a given user as features, which capture whom they listen to (information source). First, we identify the top 50 Twitter accounts followed by each user group and obtain their union, which results in 95 Twitter accounts. Then, for each user, we check whether one follows each of them and create a 95-dimensional vector as the following features.

\begin{table*}[tbh!]
\centering
\scalebox{0.85}{
\begin{tabular}{l|c|l|c||cc||cc}
\toprule
\multirow{2}{*}{\bf Model} & \multirow{2}{*}{\bf \#} & \multirow{2}{*}{\bf Features} & \multirow{2}{*}{\bf Dim.}  & \multicolumn{2}{c||}{\bf Hateful User} & \multicolumn{2}{c}{\bf High-level Hateful User} \\
 &  &  & & {\bf Macro $\boldsymbol{F_1}$} & {\bf Accuracy}  & {\bf Macro $\boldsymbol{F_1}$} & {\bf Accuracy} \\

\midrule
\emph{Baselines}
 & 1  & Majority class   &     &  36.15 & 56.62 & 41.54 & 71.05 \\
\midrule
$A$. \emph{Content} & 2 & Twitter Statistics & 6 & 67.87 & 68.58 & 46.88 & 67.37 \\
\emph{  -agnostic} & 3 &   Profile Description: SBERT & 768  & 64.60 & 66.05 & 46.88 & 67.37 \\
\emph{  features}  & 4 &  Following & 95  & 80.04 & 81.01 & 51.41 & 63.95 \\ \cmidrule{2-8}
& 5 & Twitter Stat. $+$ Prof. & 774  & 72.52 & 73.30 & 49.29 & 68.68 \\
& 6 & Twitter Stat. $+$ Fol. & 101   & 80.63 & 81.13 & \bf 58.80 & \bf 69.21 \\
& 7 & Twitter Prof. $+$ Fol. & 863  & 79.22 & 80.21 & 50.29 & 67.11\\
& 8 &  Twitter Stat. $+$ Prof. $+$ Fol. & 869  & \bf 80.98 & \bf 81.70  & 51.36 & 67.37\\ \midrule

 $B$. \emph{Content} & 9  & Shared News media & 14   & 67.87 & 71.35 & 49.82 & 68.68 \\
\emph{features} & 10  & Tweets: NELA & 85  & 74.20 & 74.68 & 54.44 & 66.32\\
& 11  & Tweets: LIWC & 73   & 76.36 & 76.75 & 55.06 & 67.63  \\
& 12  & Tweets: SBERT & 768   & 81.55 & 81.93 & 49.89 & 64.47 \\ \cmidrule{2-8}

& 13 & Media $+$ NELA & 99  & 75.38 & 76.06 & 55.98 & 68.42\\
& 14 &  Media $+$ LIWC & 87  &  78.48 & 78.83 & 56.32 & 68.42\\
& 15 & Media $+$  SBERT & 782  & \bf  82.10 & \bf 82.51 & 48.47 & 64.21\\
& 16 & NELA $+$ LIWC & 158 & 77.64 & 78.02 & 53.60 & 66.58\\
& 17 & NELA $+$ SBERT & 853 & 81.68 & 82.05 & 53.23 & 66.58 \\
& 18 & LIWC $+$ SBERT & 841 & 81.43 & 81.82 & 47.43 & 63.95\\
& 19 & Media $+$  NELA $+$ LIWC & 172 & 78.62 & 79.17 & \bf 55.98 & \bf 68.42  \\
& 20 & Media $+$  NELA $+$ SBERT & 867 & 81.95 & 82.39 & 50.12 & 66.84 \\
& 21 & Media $+$ LIWC $+$  SBERT & 855 & 81.17 & 81.59 & 51.33 & 66.05 \\
& 22 & NELA $+$ LIWC $+$ SBERT & 926 & 81.24 & 81.70& 53.78 & 67.89 \\
& 23 & Tweets: ALL & 940  & 81.49 & 81.93 & 47.08 & 62.63 \\ \midrule
\emph{Combinations}   & 24 & (Task 1) A+B: rows 8 \& 15 & 1651 & 84.03 & 84.58 & - & - \\
&  25 &  (Task 2) A+B: rows 6 \& 19 & 273 & - & - & \bf 54.58 & \bf 66.05 \\
 & 26 &  All features & 1833  & \bf 84.57 & \bf 85.04 & 49.17 & 66.05 \\
\bottomrule
\end{tabular}}
\caption{Ablation study of the proposed features for hateful user prediction (Task 1) and high-level hateful user prediction (Task 2). The Dim. column indicates the number of features (dimensions) used for each experiment.\label{tab:results_hate_vs_ref}}
\end{table*}

\subsection{Experimental Setup}
We evaluate (i) content and (ii) content-agnostic features separately and in combinations. 
We train XGBoost classifier for predicting whether or not a user will express hate towards Asians in the future. 
Although we could have adopted other models to improve the performance, our focus is on understanding risk factors rather than building the best prediction models. We thus chose XGBoost, which is highly robust across different data and problems, regularly outperforms more sophisticated models.
We perform an incremental ablation study by combining the best features from (i) and (ii) to achieve even better results. 
We split data in 80:20 ratio for training and testing.
With training data, we train and evaluate an XGBoost model using different features and feature combinations. At each iteration of the 5-fold cross-validation, we perform a grid search to tune the hyper-parameters of our XGBoost model, which are the maximum tree depth and the minimum child weight that controls for complexity and conservativeness, respectively. We use the learning rate of 0.1, gamma of 0.1, and col tree of 0.8. 
In the search process, we optimize for macro-average $F_1$ score, i.e., averaging over the classes, since our dataset is not balanced for both tasks. 
Finally, we evaluate the model on the unseen testing data. We report both macro $F_1$ score and accuracy and compare our result with the majority class baseline.


\subsection{T1: Hateful User Prediction}

Table~\ref{tab:results_hate_vs_ref} shows the evaluation results for future hate prediction grouped by feature categories. For each category, the upper rows correspond to an individual set of features, while the lower ones show their combinations. 

Rows 2-4 show that whom they follow (row 4) is the most useful feature among the content-agnostic features. As followings determine what kinds of information a user would be exposed to, this result indicates two groups are likely to be exposed to a very different set of information at least on Twitter. 

Rows 5-8 show the results of the models that combine Twitter statistics, profile description, and following features. Combining content-agnostic features generally shows improvements, except for one case---profile description features yield loss in performance when added to the following features (row 7). 
Twitter statistics and profile description improve the performance when combined with the following features, yielding the best performance among content-agnostic features (row 8). 

Rows 9-12 show that average embeddings of tweets by SBERT (row 12) work better than NELA features (row 10) or LIWC features (row 11).
We note that a combination of shared news media, NELA, and LIWC features shows improvements (row 19), but they are worse than using SBERT features. 
While news media features alone do not yield good performance, it gives a sizable improvement by +0.55 macro-F1 points (row 15) when added to the SBERT features, which is the best performance among content features. 

Finally, rows 24 and 26 show the results when combining content and content-agnostic features. The best result is achieved using all features (row 26). This combination improves over using information from contents only (row 15) by +2.47 macro-F1 points. 
The result indicates that not only users' tweets but their information sources have strong predictive power to identify those who would express hate against Asians after COVID-19, demonstrating the advantage of the user-level approach than tweet-level approach.

\subsection{T2: High-level Hateful User Prediction}

The second task is to predict whether a user would turn into high-level hateful users against Asians. 
Table~\ref{tab:results_hate_vs_ref} (Column 7 \& 8) shows the evaluation results.
We note that the dataset for this task is imbalanced (See Table~\ref{tab:dataset}), yielding high accuracy (71.05) for our baseline, majority class model. Overall, the performance of this task is not as high as hateful user prediction, reflecting the difficulty of this task.

Similar to the result of hateful user prediction, rows 2-4 suggest that `whom they follow' is more important than Twitter statistics or profile description. 
Rows 5-9 suggest that a combination of Twitter statistics and following features (row 6) result in the highest performance among content-agnostic features. 
Rows 9-12 indicate that LIWC features (row 11), which capture the psycholinguistic characteristic of users, are better than embedding and NELA features. 
Combining shared news media, NELA, and LIWC (row 19) shows a slight improvement over the model using LIWC features. 

Comparing the best model of content-agnostic features (row 6) with that of content features (row 19), unlike the results for hateful users prediction, content features perform worse than content-agnostic features. 
In other words, it is hard to distinguish high-level hateful users from low-level ones based on what they write (content). 
Instead, what information they subscribe to (following) has more explanatory power for the level of hate expression.

\subsection{Important Linguistic Features}
\label{sec:important_feature}

We examine the important linguistic features using SHAP \cite{lundberg2020local2global}.
Examining the top 20 most important features of the NELA-based model (row 10), hateful users tend to use more strong negative words and `there' while using less punctuation, positive words, and plural nouns. 
Two moral dimensions, Care/Harm and Purity/Degradation, are helpful to identify hateful users. Hateful users use more words relating to `Harm' (e.g., war, kill) and `Degradation' (e.g., disgust, gross)~\cite{curtis2011disgust}. 
Examining the LIWC based model (row 11), hateful users tend to use more `they' than `I' or `we' and use more words relating to power, risk, religion, male, wear, nonfluencies (uh, rr*) and less leisure and work related words. 
Lastly, linguistic features that predict high-level hateful users are: using more negative words and `I' and less all capitalized words, punctuation, positive and anxiety words, and internet slang.

\section{Discussion and Conclusion}

We presented a study on predicting users who would express hate towards Asians during COVID-19 by using their language use and information sources before COVID-19. 
We modeled a user by a rich set of features derived from 1) contents published by the user and 2) content-agnostic dimensions, including their Twitter statistics, profile description, and followings. 
For hateful user prediction task, our evaluation results showed that most features have a notable impact on prediction performance, which are tweets represented by SBERT, followings, LIWC, NELA, shared news media, Twitter statistics, and profile description (in this order). 
For high-level hateful user prediction, following features turn out to be more important than content features. Moreover, embedding features are worse than NELA or LIWC features, indicating that linguistic styles and information sources are crucial for predicting levels of hate towards Asians. 

Our retrospective case-control design enabled us to study the distinctive features of hateful users in comparison with reference users. We reveal their individual importance and contribution, providing interpretable insights. In contrast to previous work focusing on user features on hate content, our study sheds light on potential mechanisms and pathways (risk factors) towards online hate. In particular, our finding that one feature, following, has a strong predictive power provides compelling sociological relevance. This finding hints at social factors (which communities they belong to) potentially being dominating factor in the development of racial hatred, suggesting a strong link between social polarization and xenophobia and calling for actions of social media companies and our society.

There are some limitations to our work. 
First, this work is an observational study. Since we model information sources based on shared URLs and followings, we could not know information sources that are neither shared nor followed by users.
Second, as our study is based on the US and anti-Asian behavior, further studies would require to generalize our findings for other countries or for other ethnic minorities.
Third, our target population is those who used anti-Asian slurs, and thus, other forms of anti-Asian hate may not be included in this work. 
However, we argue that not only that the keyword-based method is a widely adopted approach for studying anti-Asian attitudes during COVID-19 ~\cite{lu2020fear,schild2020go,lyu2020sense}, but also the population using Asian slurs itself is of great importance because anti-Asian slurs 1) are unambiguously pejorative~\cite{camp2013slurring} and context-independent (Hedger2010); 2) had not been commonly used unlike other racial slurs before the pandemic~\cite{schild2020go}; and 3) have not been reclaimed by the Asian American community~\cite{croom2018asian}; and 4) their prevalence has been linked to offline behaviors and hate crimes during COVID-19~\cite{lu2020fear}. Although the set of hateful content and the set of content with anti-Asian slurs would not completely overlap, we argue that because of the aforementioned reasons, our target population (those who used slurs instead of those who posted ``hateful content'') is still a useful operationalization of anti-Asian hate. Furthermore, this operationalization allows a highly transparent and unambiguous definition of the population. If we target the population with ``hate towards Asians,'' the operationalization would inevitably require adopting methods that are much more difficult to understand and evaluate. 
While the keyword-based method can result in a skewed dataset due to the oversampling of certain keywords (e.g., n-word)~\citep{vidgen2020directions}, since we collect all tweets containing Asian slurs to study users, not tweets, we believe the sampling bias would not be a critical issue in our study. 
Furthermore, we would like to note that we exclude users who have used Asian slurs before COVID-19 to ensure that our data captures users who newly developed anti-Asian attitudes.

For future work, we plan to address the prediction task by ordinal regression that can inherently model the level of hate. 
We are also interested in characterizing hate towards Asians in other languages. 
Finally, we want to go beyond an observational study and attempt to find a potential causal relationship between information sources (biased, less credible information) and online hate.

\section*{Acknowledgements}

This research was supported by the Singapore Ministry of Education (MOE) Academic Research Fund (AcRF) Tier 1 grant, the DS Lee Foundation Fellowship awarded by the Singapore Management University, and the University of Massachusetts Lowell's COVID-19 Seed Award (Mapping Cyberhate toward Asian Americans in Digital Media Platforms in the COVID-19 Era, 2020-2022). The funders had no role in study design, data collection and analysis, decision to publish, or preparation of the manuscript.

\section*{Ethical Considerations}

While this work provides important insights on the development pathways of online hate, it also brings ethical implications by potentially being able to reveal who may turn to haters. 
We note that all data collected in this study is publicly available, and we did not attempt to use any individual-level demographic information in our study.
Yet, we are able to predict whether an individual turns into hateful after the COVID-19 at F1 = 85.57\% with simple machine learning methods based on publicly available Twitter information.
The risk of such user-based predictive tools for future offenses and misbehaviors is often underestimated and can potentially lead to algorithmic bias, as shown in the case of recidivism prediction~\cite{angwin2016machine}. 
Thus, we once again emphasize that our results should be considered as a first step for fighting online hate and further studies are required for translating it for practical use.
This work is exempt from the requirement for IRB review and approval (Reference \#11649).

While sharing tweet IDs is a common practice of studies using Twitter data, there is a risk to share tweet IDs in this work due to the sensitivity of the dataset.
For example, even if a user in the ``hateful'' user set deleted all the hate tweets, others still may be able to see whether a particular user posted anti-Asian slurs or not. 
This scenario may become possible if someone attempts to combine multiple tweet collections regarding COVID-19. 
Hence, we opt out of sharing tweet IDs. Instead, we share user-level features without revealing any personal information such as username, a profile description, etc.



\bibliographystyle{acl_natbib}
\bibliography{anthology,references_hate}

\appendix

\section{Anti-Asian Slur Words}
\label{sec:list_asian_slur_words}
We collect tweets that include any of the following slur words: chink, bugland, chankoro, chinazi, gook, insectoid, bugmen, chingchong, chee-chee, cheechee, cheena, chicom, chinaman, ching choing, chingchangchong, ching chang chong, chinki, chinky, chonky, coolie, goo-goos, googoos, gugus, huan-a, jakun,  lingling, malaun, panface, wuflu, kung flu, kungflu, yellowman, yellowwoman.

We exclude `jap,' a slur word to express hate against  Japanese, as it results in too many false-positive cases in using Twint.

\section{COVID-19 Related Keywords}
\label{sec:list_covid19_words}
We collect tweets that match one of the following keywords: coronavirus, covid, chinese virus, wuhan, ncov, sars-cov-2, koronavirus, corona, cdc, N95, epidemic, outbreak, sinophobia, china, pandemic, covd.




\end{document}